\documentstyle[twocolumn,aps]{revtex}

\begin{document}

\vskip 1cm

{\bf What can be inferred from surrogate data testing?}

In a recent Letter Palu\v{s} and Novotn\'{a} \cite{palus99}
reported statistical evidence based on surrogate data 
testing for linearity that a driven nonlinear oscillator is the 
mechanism underlying the sunspot cycle.

While their result might be true we doubt the  formal correctness 
of their conclusion. 
Surrogate data testing for linearity \cite{theiler92}
tests the null hypothesis that a linear, Gaussian, stationary,
stochastic dynamical process underlies the data, including
a possible invertible, static nonlinear observation function. 
To perform the test a feature is chosen that can capture
a violation of the null hypothesis. This feature is
evaluated for the original time series and for numerous
realizations of a process which only exhibits the linear
statistical properties of the given data. A significant
deviation of the feature evaluated for the original time 
series from the simulated distribution suggests a rejection
of the null hypothesis. The feature is usually chosen
according to a specific type of alternative hypothesis
on the underlying dynamics. In their Letter Palu\v{s} and Novotn\'{a} 
\cite{palus99} chose the amplitude-frequency correlation
as a property of nonlinear (driven) oscillators. But the rejection of the
null hypothesis based on a certain feature does not, in general, 
give evidence that the specific type of alternative that
has motivated the choice of the feature is present.
To provide evidence for a specific alternative one has to
show that the chosen feature has high power to detect the
violation by which it was motivated but no power to detect
other types of violations. Unfortunately, the null hypothesis
under consideration is such restrictive that the possible alternatives 
span a huge class of processes, see e.g.~\cite{timmer98e,timmer00a}.

With respect to the amplitude-frequency correlation considered in 
 by Palu\v{s} and Novotn\'{a}\cite{palus99} their Letter, 
for example, if the frequency of a second order linear stochastic
process is modulated with time, the resulting process analytically 
shows an amplitude-frequency correlation \cite{timmer98e}. 
A physically more plausible alternative 
hypothesis for the sun spot data arises from solar physics, see 
\cite{stix89} for review: The sun 
spots are an effect of the dynamics of the magnetic field of the sun which 
exhibits a 22 years cycle. This dynamics, a magnetohydrodynamic dynamo,
is described by a nonlinear partial differential equation which is
eventually stochastically driven.
The sun spot number represent a very specific mapping from the 
spatio-temporal magnetic field to a scalar time series. Since 
nonlinear driven partial differential equations include 
nonlinear driven oscillators
as special cases, the latter can not be distinguished from the
former based on surrogate data for the sun spots.

Summarizing, a significant amplitude-frequency correlation is
a feature of driven nonlinear oscillators, but it is not a specific
feature of these type of processes. Thus, the specific 
alternative of a driven nonlinear oscillator can not be concluded from 
a rejection of the null hypothesis.

Generally speaking, assuming that (1) no process in nature is indeed
a linear, Gaussian, stationary, stochastic dynamical one and (2) that 
one is using a feature that is capable to
detect the actual deviation, without any further information
about the process, the only thing
one can infer from surrogate data testing is whether there
are enough data for the power of the test to be large enough
to reject the by assumption untrue null hypothesis. 

\vskip 0.3cm
 J. Timmer \\
  Fakult\"at f\"ur Physik\\
 Hermann - Herder - Str. 3 \\
79104 Freiburg, Germany\\
e-mail: jeti@fdm.uni-freiburg.de

PACS: {05.45.Tp}

\end {document}